\documentclass[journal=nalefd,manuscript=article]{achemso}

\usepackage{graphicx}
\usepackage{dcolumn}
\usepackage{bm}
\usepackage{gensymb}
\usepackage{braket}
\usepackage{chemformula}
\usepackage[T1]{fontenc}

\usepackage{physics}
\usepackage{amsmath} 
\usepackage[colorlinks,allcolors=blue,urlcolor=blue]{hyperref}

\title{Realization of Hopf-link structure in phonon spectra: Symmetry guidance and High-throughput investigation}
\author{Houhao Wang}
\affiliation{National Laboratory of Solid State Microstructures and School of Physics, Nanjing University, Nanjing 210093, China}
\alsoaffiliation{Collaborative Innovation Center of Advanced Microstructures, Nanjing University, Nanjing 210093, China}
\author{Licheng Zhang}
\affiliation{National Laboratory of Solid State Microstructures and School of Physics, Nanjing University, Nanjing 210093, China}
\alsoaffiliation{Collaborative Innovation Center of Advanced Microstructures, Nanjing University, Nanjing 210093, China}
\author{Ruixi Pu}
\affiliation{National Laboratory of Solid State Microstructures and School of Physics, Nanjing University, Nanjing 210093, China}
\alsoaffiliation{Collaborative Innovation Center of Advanced Microstructures, Nanjing University, Nanjing 210093, China}

\author{Xiangang Wan}
\affiliation{National Laboratory of Solid State Microstructures and School of Physics, Nanjing University, Nanjing 210093, China}
\alsoaffiliation{Collaborative Innovation Center of Advanced Microstructures, Nanjing University, Nanjing 210093, China}
\alsoaffiliation{Jiangsu Physical Science Research Center}
\author{Feng Tang}\email{fengtang@nju.edu.cn}
\affiliation{National Laboratory of Solid State Microstructures and School of Physics, Nanjing University, Nanjing 210093, China}
\alsoaffiliation{Collaborative Innovation Center of Advanced Microstructures, Nanjing University, Nanjing 210093, China}

\begin{document}
\begin{abstract}
The realization of Hopf-link structure in the Brillouin zone is rather rare hindering the comprehensive exploration and understanding of such exotic nodal loop geometry.  Here we first tabulate 141 space groups hosting Hopf-link structure and then investigate Phonon Database at Kyoto University consisting of 10034 materials to search for phonon realization of the Hopf-link nodal structure.
It is found that almost all the investigated materials own nodal loops or nodal chains while only 113 materials can host Hopf-link structure in phonon spectra, among which 8 representative materials are manually selected to showcase relatively clean Hopf-link structure including LiGaS$_2$, LiInSe$_2$, Ca$_2$Al$_2$Si(HO$_4$)$_2$, Ca$_7$GeN$_6$, Al(HO)$_3$, NaNd(GaS$_2$)$_4$, Ga$_5$(PS)$_3$ and RbTh$_3$F$_{13}$.
The visible phonon drumhead surface states corresponding to the nodal loops in the Hopf-link structure are further demonstrated using Ga$_5$(PS)$_3$ as an example.
The listed 113 crystalline materials provide a good platform for experimentalists to further explore the interesting properties related to Hopf-link structure.
\end{abstract}
\bigskip

\begin{figure*}[!ht]
\centering\includegraphics[width=1\textwidth]{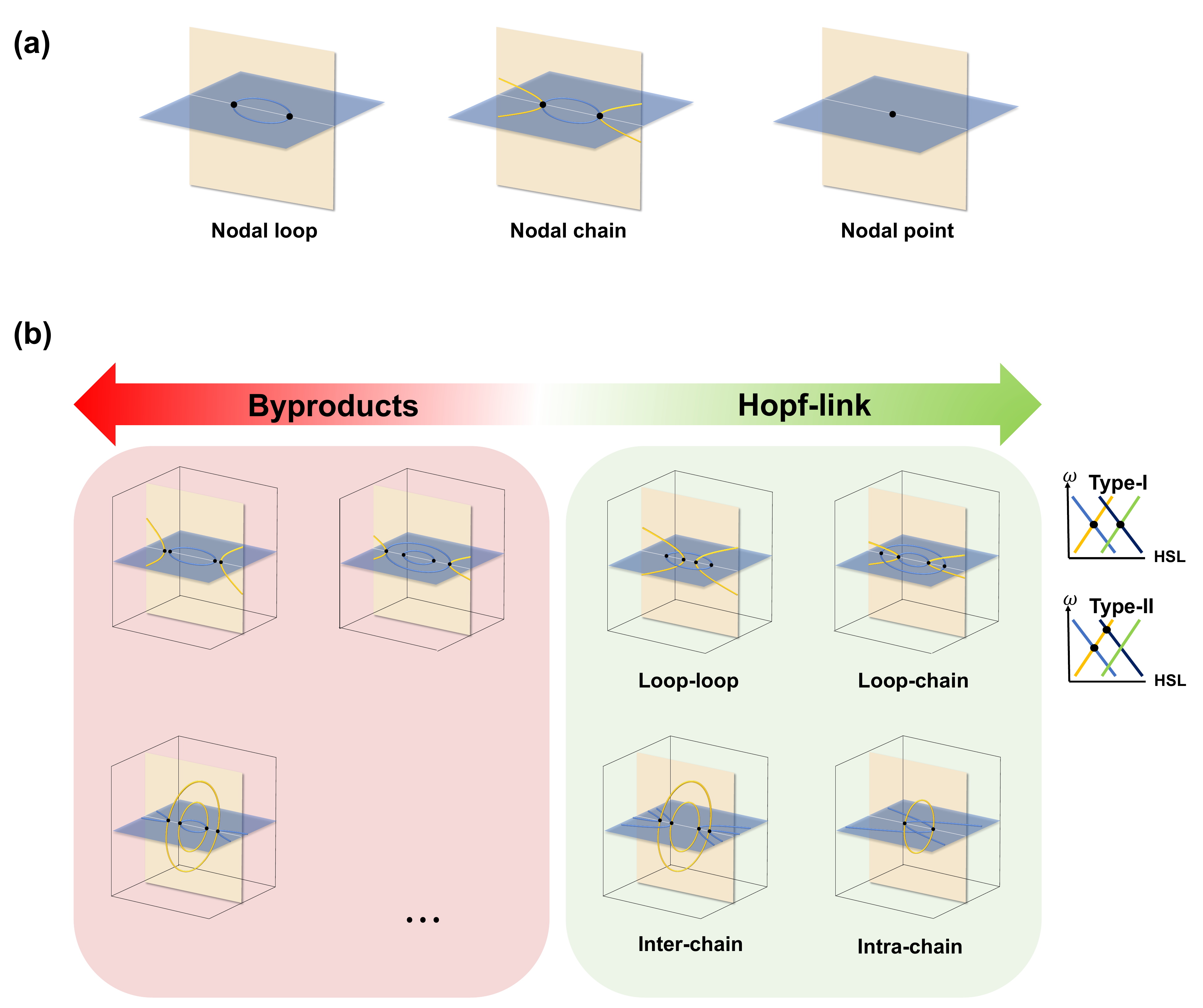}
\caption{(a) All three possible cases of BCs (indicated by black circles) enumerated using two different irreps in HSL (indicated by a white line). In the first case, the BC lies in a nodal loop within a HSPL (indicated by a blue plane). In the second case, the BC is a point of tangency between loops within two HSPLs (indicated by yellow and blue planes) or within additional HSPLs, forming a nodal chain structure. In the third case, the BC is an isolated nodal point. (b) Four kinds of Hopf-link structures are formed by nested loops/chains, and three kinds of representative byproducts come from separated loops/chains. The right panels show that type-I (II) Hopf-link structures (including loop-loop, loop-chain, and inter-chain) originate from two (or three) bands.}
\label{schematic}
\end{figure*}

The research on topological quantum states provides us with a more advanced understanding of electronic behavior in solids \cite{kane2005prl,hasanrmp2010,qi2011rmp,weyldirac2018rmp}. By analogy to electronic system, the concept of topology was extended to bosons \cite{colloquium2012rmp,yang2015prl,topo2016np,susstrunk2016pnas,liu2018ncr,ma2019nrl,liu2020afm}.
As topological semimetals for electronic band structures \cite{3ddirac-young-prl,diracreview-balatsky-ap,classDirac-nagaosa-nc,weyl2015science,dirac2015prl,du2015dirac,weyl2015nature,dirac2016prl,dirac2016np,weyldirac2018rmp,nodalline2015prb,nodalline2015prl2,nodalline2016cpb,nodallineprl2016,tomas2016nodalchain2,du2017cate,chain2017prl,chain2017nc,chain2018np,chain2018prb,chain2019prb,surface2016nanosale,surface2016prb,surface2017prb,surface2018prb,surface2018prb_2,surface2018prb_3,surface2019sadv,wulinprb2020,surface2021prb,tangfeng2022prb,wangwencheng2024prb}, various topological phonon band crossings (BCs) have been proposed such as Weyl/Dirac phonons \cite{wanprb2011, jin2018nano,xie2019prb,xia2019prl,liu2020npj,chen2021prl}, three- and six-degenerate nodal point phonons \cite{singh2018prm,xie2021prb}. Different from nodal point, high-dimensional geometric shapes in the Brillouin zone (BZ), such as straight  line \cite{li2020prb,liu2021prb}, chain \cite{zhu2022npj,chen2021njp},  box/net/cage \cite{wang2021prb,cucl2021prb,li2023fip,zhoufeng2021prb,wang2022prb,cucl2021prb,liu2022mtp}, and surface \cite{liuqingbo2021prb,xiechengwu2021prb,xiechengwu2022prb}.  Recently, Hopf-link nodal structure with two nesting nodal loops (see loop-loop in Figure \ref{schematic}b) has been proposed \cite{chen2017topological,wangzhong2017prb,Ezawa2017prb,poyaochang2017prb,changguoqing2017prl}.
Notably, the Hopf-link structure leading to stitching surface states proposed  in the ferromagnetic full Heusler compound Co$_2$MnGa in Ref. \cite{changguoqing2017prl}, been observed experimentally \cite{hopf2022nature}.
To date, the proposed realization for the Hopf-link structure in  phonon spectrum usually needs a symmetry-breaking perturbation inducing a finite gap to a nontrivial nodal point \cite{liu2022nl,RTP-NC}. However, the resulting Hopf-link is restricted within a small volume. Very few realistic crystalline materials with extended Hopf-link structure thus hinders exploration of the predicted properties such as Seifert boundary
states \cite{hopf2022nature} and  unique magnetotransport properties \cite{zhou2018hopf,xieyuee2019prb} and the promising applications \cite{photon2020prl}.

Space groups (SGs) have been shown to lead to various topological phases, and high-throughput screening has been applied to discover and categorize thousands of topological materials based on symmetry analysis
\cite{hpc2019nature2,hpc2019nature,hpc2019nature3,lijiangxu2021nc,tangfeng2021prb,hpc2022science,xu2022catalogue}.
The topological phonon database built by Xu et al. \cite{hpc2019nature2} only calculated phonon irreducible representations (irreps) at high-symmetry points.

\begin{figure*}[!h]
	\centering\includegraphics[width=1. \textwidth]{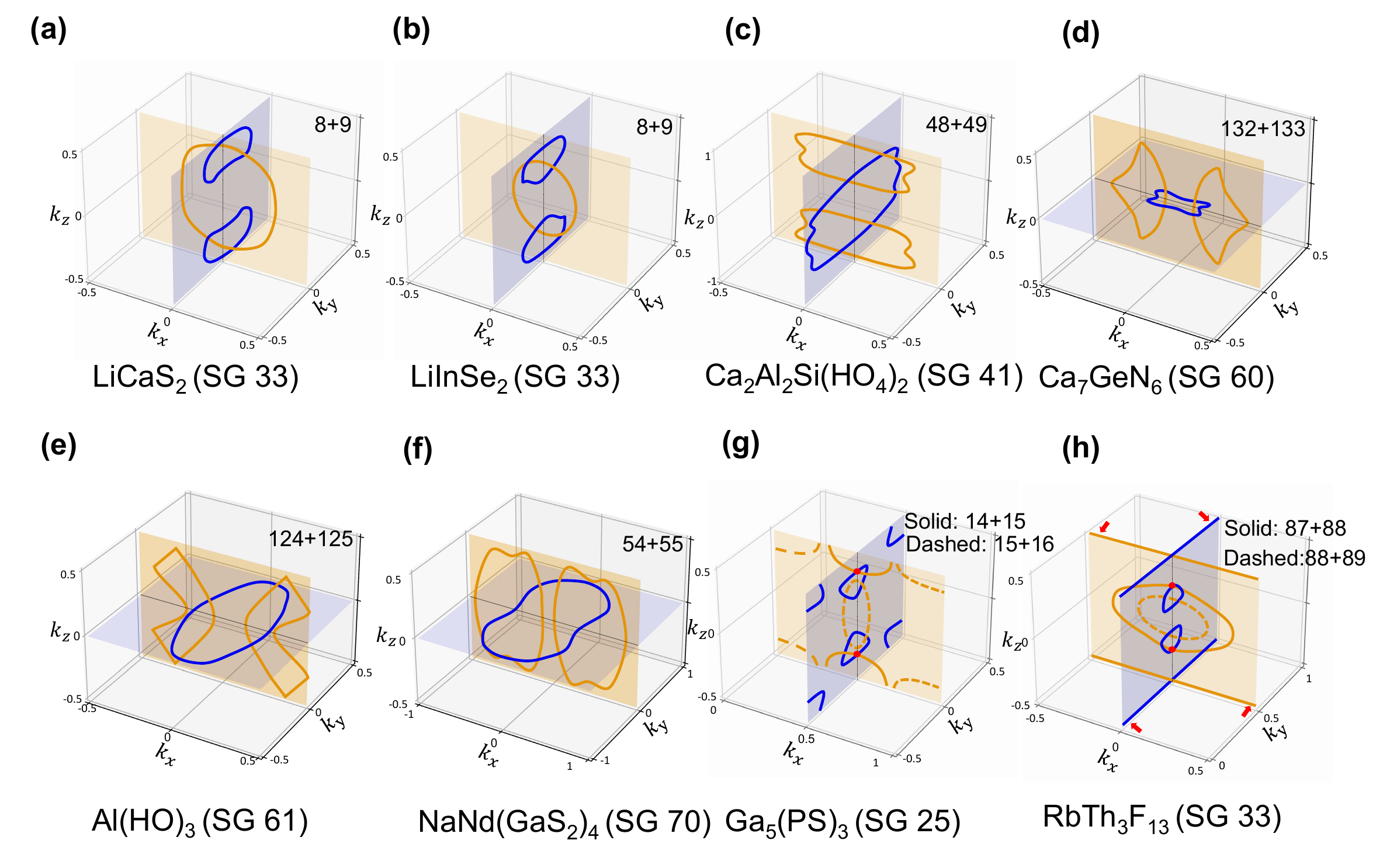}
   \caption{The Hopf-link structures in 8 representative materials. (a-f) showcase Type-I loop-loop structures while (g-h) showcase Type-II loop-chain structures. For each material, we provide  the chemical formula, SG number (in the parenthesis after the chemical formula) and the branches forming each nodal loop (e.g., 8+9 in (a) means that the sketched nodal loops are formed from the 8th and 9th branches). Note that in (g) and (h), we show Type-II loop-chain structures (in which the nested loops and chains are formed by two sets of branch doublets. e.g., one nodal chain formed by the 14th and 15th branches is nested with another nodal loop formed by the 15th and 16th branch, as shown in (g)). The yellow/blue curves represent nodal structure lying in light yellow/light blue planes respecting HSPLs, respectively. The HSPLs in (a)-(c) are $k_x=0$ and $k_y=0$; those in (d)-(f) are $k_y=0$ and $k_z=0$; those in (g) are $k_x=\frac{1}{2}$ and $k_y=0$; and those in (h) are $k_x=0$ and $k_y=\frac{1}{2}$.  In (g) and (h), nodal chains touch at the red circles. The straight nodal lines in (h) (indicated by red arrows) coexist with the Hopf-link structure. The units of $k_x,k_y,k_z$ are $\frac{2\pi}{a}$,$\frac{2\pi}{b}$,$\frac{2\pi}{c}$, respectively, where $a,b,c$ are lattice parameters for each material.}
\label{hopf-link}
\end{figure*}

In this work, by completely classifying all BCs (formed by different irreps) lying in high-symmetry lines (HSLs) for 230 SGs based on compatibility relations (CRs) in phonon spectra, we first tabulate 141 SGs that allow Hopf-link structures composed of two sets of nodal chain (loop) that are nested with each other in Table \ref{table}. Then we perform a high-throughput investigation to discover Hopf-link structure based on Phonon Database at Kyoto University (PhononDB@kyoto-u) \cite{DB} with first-principles computed force constants for 10034 crystalline materials.
Based on our calculation results, we find almost all the investigated materials exhibit loops and chains that are separated, but only 113 materials can host Hopf-link structure.
We further manually select 8 representative materials crystallized in SGs 25, 26, 33, 60, 61 (with a primitive orthorhombic lattice), SG 41 (with a base-centered orthorhombic lattice) and SG 70 (with a face-centered orthorhombic lattice) as examples to showcase relatively high-quality Hopf-link structure, as shown in Figure \ref{hopf-link}(a-h).
In addition, we demonstrate that visible phonon drumhead surface states arising from nodal loops in Hopf-link structure, using Ga$_5$(PS)$_3$ as an example.

We first briefly overview the methodology of identifying  Hopf-link structure based on BCs of HSLs.
All possible types of BCs can be enumerated using two different irreps (denoted by $D_G$ and $D'_G$) of the little group associated with HSL G. The CRs can indicate the nearby nodal structure around the BC by examining all neighboring high-symmetry planes (HSPLs) containing the HSL: that the BC lies in one nodal loop within some neighboring HSPL if and only if the corresponding CRs are $D_G\rightarrow D, D'_G\rightarrow D'$ where $D,D'$ represent two different irreps of this HSPL, as illustrated in the first case of Figure \ref{schematic}a.
Note that if two or more such HSPLs can be found, the BC is then a touching point of two or more nodal loops within respective HSPLs, forming a nodal chain structure, as illustrated in the second case of Figure \ref{schematic}a. Once such neighboring HSPL cannot be found, the BC is then an isolated nodal point, and there is no emanating nodal loop threading the BC, as illustrated in the third case of Figure \ref{schematic}a.
For BCs lying in loops or chains, four kinds of Hopf-link structures can be formed: loop-loop, loop-chain, inter-chain, and intra-chain, as shown in the green area of Figure \ref{schematic}b.
A loop-loop structure requires identifying BCs that lie loops of neighboring HSPLs: the HSPLs that accommodate the two loops are distinct, while also ensuring that loops are nested within each other.
A loop-chain structure requires identifying two types of BCs: one lies in a loop and the other in a chain, while ensuring the loop is nested with the chain.
If two chains are nested together, an inter-chain structure can be formed.
In fact, because of time-reversal symmetry and the periodicity of the BZ, a single chain (intra-chain in Figure \ref{schematic}b) can also be considered as a nested geometric structure.
Based on CRs, we enumerate 141 SGs (in orthorhombic, tetragonal, trigonal, hexagonal and cubic crystal systems) that allow for the diagnosis of the above four kinds of Hopf-link structures with time-reversal symmetry in the spinless setting suitable for phonons utilizing BCs within HSLs, as shown in Table \ref{table}.
The details of results of BCs within HSLs and positions possibly hosting each kind of Hopf-link geometric structure are provided in the Supporting Information (SI) Section S5.
For loop-loop, loop-chain and inter-chain geometric structures, we further focus on two band types shown in the right panel of Figure \ref{hopf-link}b: Type-I Hopf-link structure originates from loops or chains formed by one band doublet, while Type-II Hopf-link structure originates from loops or chains formed by two different doublets.

\begin{figure*}[!h]
	\includegraphics[width=1 \textwidth]{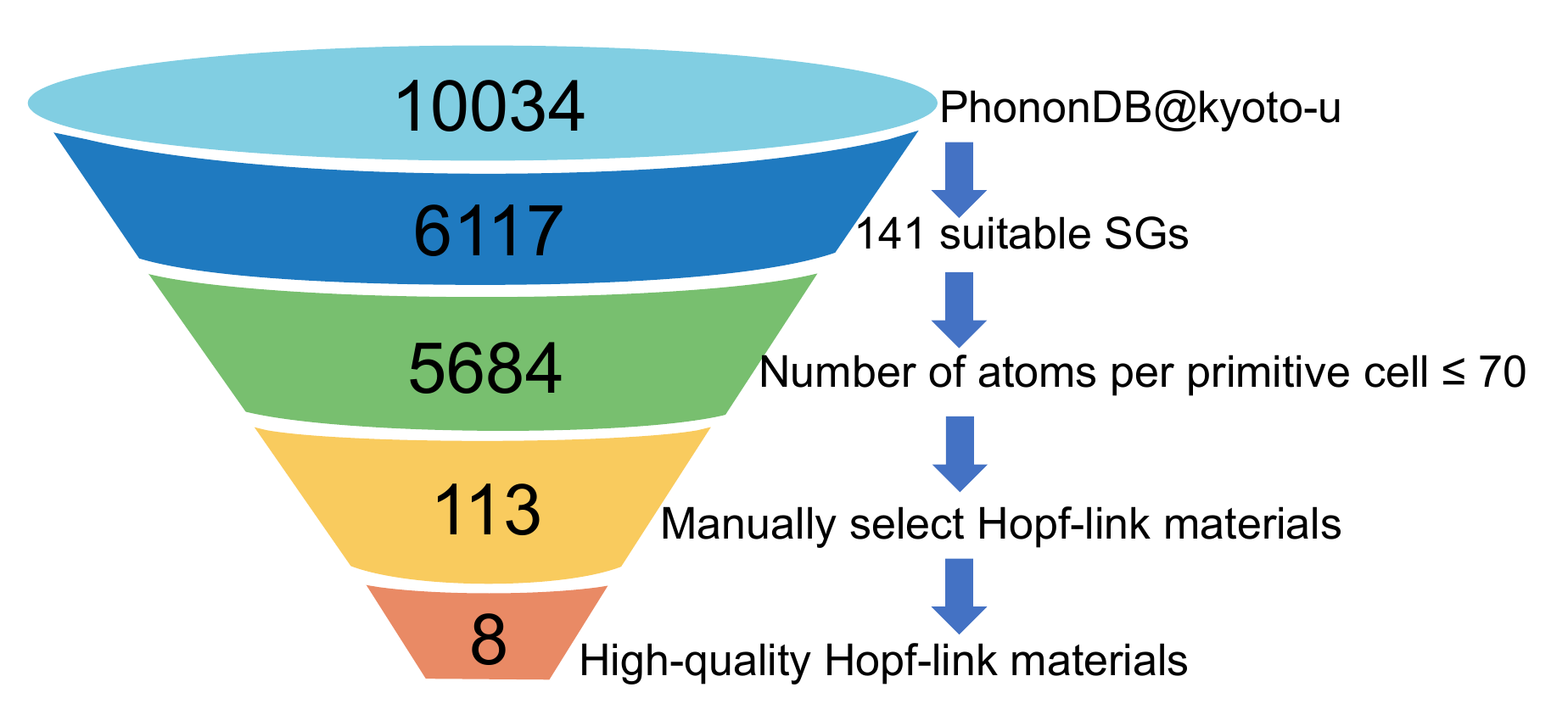}
   \caption{Workflow of material screening for Hopf-link phonons. In the first step, two filters on SGs and number of atoms are applied: 6117 materials in PhononDB@kyoto-u \cite{DB} are crystallized in any of the 141 SGs in Table \ref{table} allowing for Hopf-link phonon by calculating irreps in HSL, and out of these materials, 5684 materials are further selected owning the number of atoms per primitive unit cell (N$_a$) $\leq$ 70. We then perform irrep calculations of phonon wavefunctions for these materials along HSLs to identify BCs composed of different irreps, by which many loops and chains threading the BCs can be identified. In the last step, by manually checking whether loops/chains are nested with each other, we finally collect 113 candidate materials hosting Hopf-link structure in the phonon spectrum. We also further sort out 8 representative materials (Figure \ref{hopf-link}(a-h)).}
\label{workflow}
\end{figure*}

Next, we introduce the workflow for material screening of Hopf-link structures in phonon spectra (Figure \ref{workflow}). 
Out of 10034 materials with first-principles calculated force constants in  Kyoto University (PhononDB@kyoto-u) \cite{DB}, 6117 are crystallized in any of 141 SGs.
We further restrict the number of atoms per primitive unit cell to be $\leq$ 70, resulting in a total of 5684 materials.
Then we obtain the BCs for the entire frequency range by calculating the irreps of phonon wavefunctions within HSLs.
Next, we quickly diagnose Type-I/II loop-loop, Type-I/II loop-chain, Type-I/II inter-chain and intra-chain structures, respectively.
Here, we provide a brief overview of our strategy for searching for Type-II loop-chain structures (see details in SI section S4).
The Type-II loop-chain structure requires that nested loop and chain originate from three band branches (band indices: $n-n+2$), with detailed positions of loops and chains available in the SI section S5. 
The screening of the Type-II loop-chain structure is divided into three steps:
first, extract BCs from bands (band indices: $n$, $n+1$), indicated by BC1, and extract BCs from bands (band indices: $n+1$, $n+2$), indicated by BC2; second, ensure that BC1 (or BC2) corresponds exclusively to one loop or more loops while ensuring that BC2 (or BC1) contains one chain or more chains; third, check whether loops and chains are nested with each other in the BZ.

Based on above workflow, we find that 113 materials include Type-I/II loop-loop, Type-I/II loop-chain and Type-II inter-chain structure (see details in SI Section S4).
Furthermore, we also find many byproducts, such as
separated loops and chains,  as illustrated in the red area of Figure \ref{hopf-link}b.
Then, we manually select 8 representative materials with relatively clean Hopf-link structures, including LiGaS$_2$ \cite{cailiao1}, LiInSe$_2$ \cite{cailiao2}, Ca$_2$Al$_2$Si(HO$_4$)$_2$,
Ca$_7$GeN$_6$  \cite{cailiao5}, Al(HO)$_3$, NaNd(GaS$_2$)$_4$ \cite{cailiao6}, Ga$_5$(PS)$_3$ and RbTh$_3$F$_{13}$ \cite{cailiao7} (Figure \ref{hopf-link}(a-h)).
The Hopf-link structures in Figure \ref{hopf-link}(a-f) consist only of loops, whereas the Hopf-link structures in Figure \ref{hopf-link}(g-h) contain both loops and chains. Furthermore, coexisting straight nodal lines can be observed in Figure \ref{hopf-link}(h).
Next, we select Ga$_5$(PS)$_3$ as a representative example to elucidate the Hopf-link structure in detail.
We provide data on the formation of various nodal loop configurations of the other seven materials in SI Section S3.

\begin{table}[!t]
\caption{All HSLs allowing for diagnosing phonon Hopf-link structures (Figure \ref{schematic}). SGs hosting loop-loop, loop-chain, inter-chain and intra-chain structures are printed in bold, while SGs only hosting inter-chain are printed in regular font.}
\resizebox{0.5\textwidth}{!}{
\begin{tabular}{cc}
\hline
\hline
HSL &  SGs \\\hline
A&\textbf{38}, \textbf{39}, \textbf{47}, \textbf{50}, \textbf{51}, \textbf{55}, \textbf{59}, \textbf{65}, \textbf{67}, \textbf{69}\\
B&\textbf{47}, \textbf{50}, \textbf{51}, \textbf{55}, \textbf{59}, \textbf{65}, \textbf{67}, \textbf{69}\\
C&\textbf{47}, \textbf{49}, \textbf{51}, \textbf{53}, \textbf{54}\\
D&\textbf{47}, \textbf{49}, \textbf{57}, \textbf{217}, \textbf{229}\\
DT&\textbf{47-74}, \textbf{123-142}, 156-159, 162-165, 183-194, \textbf{200-206}, 215-230\\
E&\textbf{47}, \textbf{48}, \textbf{51}\\
G&\textbf{25-27}, \textbf{30}, \textbf{47}, \textbf{49}, \textbf{57}, \textbf{229}\\
H&\textbf{25-29}, \textbf{31}, \textbf{35-37}, \textbf{42}, \textbf{47}, \textbf{49}, \textbf{51}, \textbf{53}, \textbf{54}, \textbf{63-69}\\
LD&\textbf{25-37}, \textbf{42-74}, 99-142, 160, 161, 166, 167, \textbf{189-194}, 215-230\\
P&\textbf{47}, \textbf{48}, \textbf{52}, 157, 159, 162, 163, 183-186, 191-194\\
PA&157, 159, 189, 190\\
Q&\textbf{25-27}, \textbf{32-34}, \textbf{47-50}, \textbf{189}, \textbf{191}\\
R&\textbf{187}, \textbf{191}\\
S&\textbf{123}, \textbf{125}, \textbf{127}, \textbf{129}, \textbf{132}, \textbf{134}, \textbf{136}, \textbf{138}, \textbf{221}, \textbf{224}\\
SM&\textbf{38-41}, \textbf{47-74}, \textbf{123-142}, \textbf{187}, \textbf{188}, \textbf{191-194}, \textbf{221-230}\\
T&\textbf{123}, \textbf{126}, \textbf{131}, \textbf{134}, \textbf{200}, \textbf{201}, 215, 218, 221-224\\
U&\textbf{123}, \textbf{125}, \textbf{127}, \textbf{129}, \textbf{131}, \textbf{133}, \textbf{135}, \textbf{137}, \textbf{183-186}, \textbf{191-194}\\
V&99-106, 111, 112, 115-118, 123-126, 131-134, \textbf{202}, \textbf{225}, \textbf{226}\\
W&\textbf{99}, \textbf{101}, \textbf{103}, \textbf{105}, \textbf{107}, \textbf{108}, \textbf{115}, \textbf{116}, \textbf{121}, \textbf{123}, \textbf{124}, \textbf{131}, \textbf{132}, \textbf{139}, \textbf{140}\\
Y&\textbf{123}, \textbf{124}, \textbf{131}, \textbf{132}, \textbf{139}, \textbf{140}\\
Z&\textbf{200}, \textbf{221}, \textbf{223}\\
ZA&\textbf{200}\\
\hline
\hline
\end{tabular}}
\label{table}
\end{table}

\begin{figure*}[!h]
	\centering\includegraphics[width=1. \textwidth]{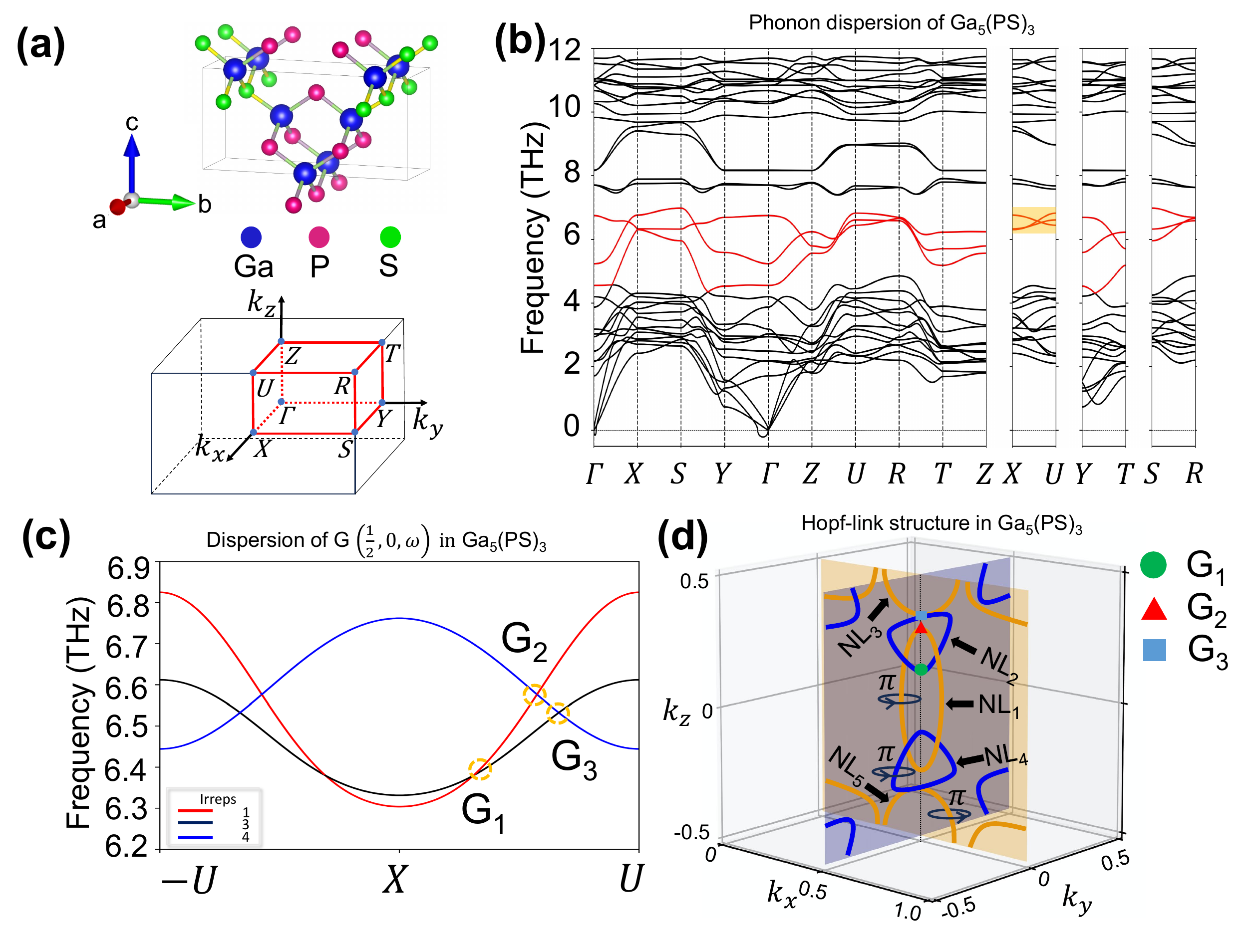}
\caption{(a) The crystal structure and the bulk BZ of Ga$_5$(PS)$_3$ (SG 25). (b) The phonon dispersion of Ga$_5$(PS)$_3$. The Hopf-link structure arises from bands with indices 14-16, emphasized by solid red lines. (c) The phonon dispersion of 14th, 15th and 16th bands along the HSL G $(\frac{1}{2},0,w)$ ($-U-X-U$) path in the BZ. The color of the lines represents irreps to which bands belong. The irreps are indicated by colors of bands. Note that there are three BCs labeled by the dashed yellow circles, named G$_1$, G$_2$ and G$_3$, with irreps of $1\oplus3$, $1\oplus4$ and $3\oplus4$, respectively. (d) The Hopf-link structure in the 3D BZ. G$_1$ (marked with green dot) lies in a blue nodal loop within HSPL $k_x=\frac{1}{2}$ (indicated by NL$_2$), while G$_3$ (blue square) lies in the nodal chain composed of NL$_2$ and one yellow nodal loop within $k_y=0$ plane (NL$_3$). G$_2$ (red triangle) lies in a yellow nodal loop within $k_y=0$ plane (NL$_1$), which is nested with NL$_2$. NL$_4$ and NL$_5$ can be related to NL$_2$ and NL$_3$, respectively, through time-reversal symmetry. All NL$_i$ (where i=1,2,3,4 and 5) all have a $\pi$ value of Berry phase.}
\label{spectrum}
\end{figure*}

Type-II loop-chain structure candidate material Ga$_5$(PS)$_3$ is crystallized in SG 25. The crystal structure and the BZ are shown in Figure \ref{spectrum}(a). The primitive cell contains 11 atoms: 5 Ga, 3 P and 3 S atoms. We showcase first-principles calculation results of Ga$_5$(PS)$_3$ in \ref{spectrum}(b). The phonon dispersion shows only slight imaginary frequencies near $\Gamma(0,0,0)$, indicating that Ga$_5$(PS)$_3$ is dynamically stable.
The BCs of bands (band indices: 14-16) which are emphasized with solid red lines along HSL G $(\frac{1}{2}, 0, w)$ ($X-U$) successfully diagnose Type-II loop-chain structure by our strategy.
As shown in Fig. \ref{spectrum}(c), the BCs in HSL G, denoted by the yellow dashed circle, are labeled as G$_i$ (for $i=1, 2, 3$) respectively. The coordinates for G$_1$, G$_2$, G$_3$ are $(0.5, 0, 0.154)$, $(0.5, 0, 0.288)$ and $(0.5, 0, 0.333)$, with corresponding frequencies of 6.379 THz, 6.575 THz and 6.532 THz, respectively.
And we find that G$_1$, G$_2$ and G$_3$ are associated with irreps of $1\oplus3$, $1\oplus4$ and $3\oplus4$, respectively.
It is possible to form Type-II loop-chain structure: G$_1$ and G$_2$ must lie in a nodal loop within HSPL $k_x=\frac{1}{2}$ and $k_y=0$, respectively, while G$_3$ must lie in a nodal chain based on CRs.
By checking nodal structure geometry in 3D BZ, as shown in Fig. \ref{spectrum}(d), we find that G$_i$ (with $i=1, 2, 3$) utilizes Type-II loop-chain structure: G$_1$ lies in the blue nodal loop within HSPL $k_x=\frac{1}{2}$ (indicated by NL$_2$) while G$_3$ lies in the nodal chain composed of NL$_2$ and the yellow nodal loop within HSPL $k_y=0$ (indicated by NL$_3$). G$_2$ lies in one yellow nodal loop within HSPL $k_y=0$ (indicated by NL$_1$), which is nested with NL$_2$. Furthermore, we use the constructed $k \cdot p$ model to fit the first-principles phonon spectra and sketch loops and chains using the concrete $k \cdot p$ model (see details in SI Section S2).
It is worth mentioning that predicted nodal structures based on CRs are further confirmed by our effective $k \cdot p$ model, which could be used in future studies of effects on the behavior of these BCs and the associated nodal loops by external fields.

\begin{figure*}[!tbp]
	\centering\includegraphics[width=1. \textwidth]{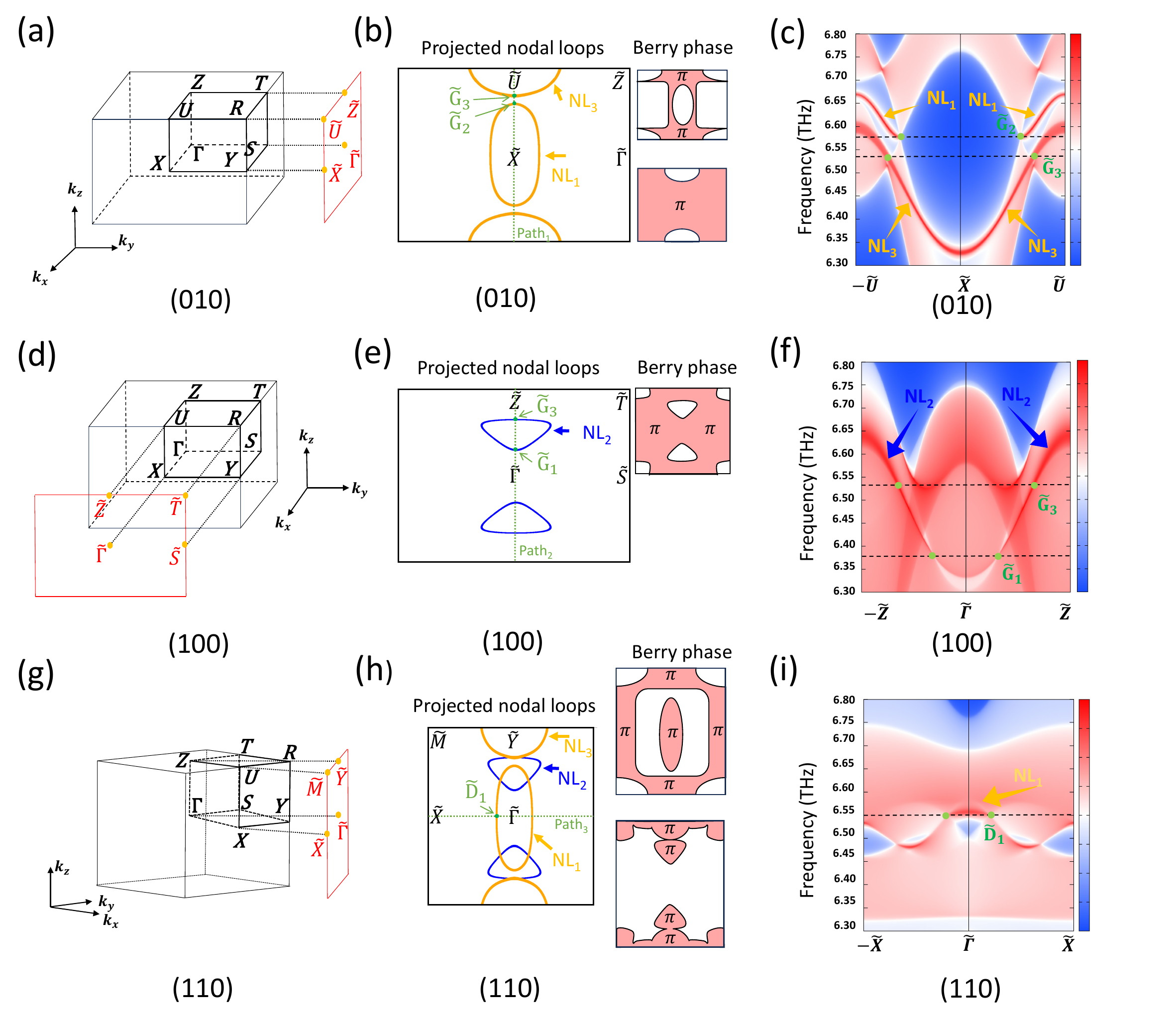}
   \caption{
   (a) The 3D BZ of SG 25 and the (010) surface BZ. (b) The (010) surface projections of the NL$_1$ and NL$_3$ (see Fig. \ref{spectrum} (i)). The right panel shows Berry phase values (either 0 or $\pi$) for paths normal to the surface by setting the occupied bands to 15th (corresponding to NL$_1$) and the 14th (corresponding to NL$_3$), respectively. (c) The LDOS along path (indicated as path$_1$ in (b)) on the (010) surface BZ, showing drumhead surface states protected arising from NL$_1$ and NL$_3$ protected by Berry phase $\pi$ (denoted by yellow arrows). (d) The (100) surface BZ. (e) Surface projections of NL$_2$ on the (100) surface BZ, with Berry phase values for normal paths.  (f) The LDOS along path$_2$ on the (100) surface BZ, showing drumhead surface states arising from NL$_2$ protected by Berry phase $\pi$ (blue arrow). (g) The (110) surface BZ. (h) The (110) surface projections of the NL$_i$ (with $i=1,2,3$) (see Fig. \ref{spectrum} (i)). The right panel shows Berry phase values for normal paths by setting the occupied bands to 15th (corresponding to NL$_1$) and 14th (corresponding to NL$_2$ and NL$_3$), respectively. (i) The LDOS along path$_3$ on the (110) surface BZ, showing drumhead states arising from NL$_1$ protected by Berry phase $\pi$ (yellow arrow).}
\label{surface}
\end{figure*}

In the following, we demonstrate the surface states related to the Hopf-link structure in Ga$_5$(PS)$_3$. All loops composing the Hopf-link structure in the $6.30 \sim 6.80$ THz range, namely, the yellow nodal loops NL$_1$ and NL$_3$ in $k_y=0$ plane and the blue nodal loop NL$_2$ in $k_x=\frac{1}{2}$ plane, are found to carry non-vanishing Berry phases, as shown in Figure \ref{spectrum}(d). It is expected that surface drumhead states will be present, and  we choose the (010), (100), and (110) surfaces to showcase them.
For the (010) surface (Figure \ref{surface}a), which is parallel to HSPL $k_y=0$ plane, the corresponding projections of the NL$_1$ and NL$_3$ are shown in Figure \ref{surface}b. 
The red zone (in right panel of Figure \ref{surface}b) represents $\pi$ value of Berry phase for paths normal to the (010) surface corresponding NL$_1$ and NL$_3$, are expected to possess the topologically protected drumhead surface states. We plot the surface local density of states (LDOS) in Figure \ref{surface}c.
It can be observed that, as indicated by the yellow arrows in Figure \ref{surface}c, the topologically protected drumhead surface states of NL$_1$ and NL$_3$ can be clearly identified.
With respect to the (100) surface parallel to HSPL $k_x=\frac{1}{2}$, the topologically protected surface states indicated by the blue arrows, which originated from NL$_2$, can be found, as shown in Figure \ref{surface}(d-f).
Furthermore, on the (110) surface (Figure \ref{surface}g), which includes all projected loops (Figure \ref{surface}h), we find a flatter topologically protected drumhead surface state with a large density of states from NL$_1$ (Figure \ref{surface}i), which might be important for surface-phonon-related physics, such as low-dimensional superconductivity \cite{liu2022nl}.

Experimentally, with the advancement of phonon measurement techniques, some ideal material candidates featuring topological phonons have been gradually measured and confirmed.
The high energy resolution inelastic x-ray scattering (HEIXS) \cite{IXS1,IXS2} can observe the BCs of phonon spectra along specific momentum directions.
The double-Weyl phonon in FeSi \cite{miao2018prl} and the phononic nodal line in MoB$_2$ \cite{zhangtt2019prl} have been observed using HEIXS.
Moreover, the recently developed 2D high resolution electron energy loss spectroscopy (HREELS) \cite{HREELS1,HREELS2} has achieved effective mapping of phonon dispersion across the entire momentum space and is a suitable tool for measuring phonon surface states. Recently, Li et al. \cite{lijiade2023prl} directly observed  nodal ring phonons and Dirac phonons in graphene using HREELS.
The 113 realistic materials hosting phonon Hopf-link structures, some of which have been synthesized, such as infrared nonlinear optical crystals LiCaS$_2$ \cite{cailiao1} and LiInSe$_2$ \cite{cailiao2}, could attract immediate experimental interest in the future.

In summary, we carried out a high-throughput calculation on crystalline materials in PhononDB@kyoto-u \cite{DB} to identify candidates for realization of the Hopf-link structure. We find 113 pristine crystalline materials that exhibit Hopf-link structures, which can provide a platform to assist experimentalists in further exploring their interesting properties with practical application value.
Our work also demonstrates that computing irreps in HSLs is an efficient strategy for searching for Hopf-link structures, which is expected to be applied to other systems in the future, such as electronic band structures in non-magnetic and magnetic materials, as well as other types of bosons in photons or magnons.

\section{ACKNOWLEDGMENTS}
This paper was supported by the National Natural Science Foundation of China (NSFC) under Grants No. 12188101, No. 12322404, No. 12104215, No. 11834006, the National Key R\&D Program of China (Grant No. 2022YFA1403601), Innovation Program for Quantum Science and Technology, No. 2021ZD0301902 and Natural Science Foundation of Jiangsu Province (No. BK20233001, BK20243011). F.T. was also supported by the Young Elite Scientists Sponsorship Program by the China Association for Science and Technology. X.W. also acknowledges support from the Tencent Foundation through the XPLORER PRIZE.

\clearpage
\bibliography{Ref}

\end{document}